# Imaging Nanophotonic Modes of Microresonators using a Focused Ion Beam

Kevin A. Twedt[a,b]†, Jie Zou[a,b]†, Marcelo Davanco[a], Kartik Srinivasan[a], Jabez J. McClelland[a], Vladimir A. Aksyuk[a]*

[a]Center for Nanoscale Science and Technology, National Institute of Standards and Technology, Gaithersburg, MD 20899, USA
[b]Maryland Nanocenter, University of Maryland, College Park, MD 20742, USA

†These authors contributed equally to this work.
*vladimir.aksyuk@nist.gov

**Optical microresonators have proven powerful in a wide range of applications, including cavity quantum electrodynamics[1–3], biosensing[4], microfludics[5], and cavity optomechanics[6–8]. Their performance depends critically on the exact distribution of optical energy, confined and shaped by the nanoscale device geometry. Near-field optical probes[9] can image this distribution, but the physical probe necessarily perturbs the near field, which is particularly problematic for sensitive high quality factor resonances[10,11]. We present a new approach to mapping nanophotonic modes that uses a controllably small and local optomechanical perturbation introduced by a focused lithium ion beam[12]. An ion beam (radius ≈50 nm) induces a picometer-scale dynamic deformation of the resonator surface, which we detect through a shift in the optical resonance wavelength. We map five modes of a silicon microdisk resonator (Q≥20,000) with both high spatial and spectral resolution. Our technique also enables in-situ observation of ion implantation damage and relaxation dynamics in a silicon lattice[13,14].**

Optical microresonators have a long and successful history as ultrasensitive detectors, mainly thanks to enhanced light-matter interactions resulting from their appreciable photon lifetime and optical field confinement. The numerous applications typically involve detecting small changes in the optical resonances when a perturber, such as a nanomechanical resonator[6–8], biomolecule[5], or single atom[15], moves within the near-field of the device.  Knowledge of the optical mode field distribution is thus of critical importance to predicting the interaction with the perturber, and optimizing device performance.  But the ultrahigh sensitivity also makes it challenging to measure the mode shape directly and non-invasively, especially as the device dimensions continue to decrease[8].

Near-field scanning optical microscopy (NSOM) has been the most widely used tool for mapping nanoscale fields.  While NSOM has been successfully applied to measure the evanescent field near the surfaces of microresonators[10,11], it is hampered by the often-strong interaction between the tip and the optical mode being measured, which can lead to a degraded quality factor (Q) and shifts in the resonance wavelength.  Reducing the disturbance by minimizing the probe size and moving it farther away from the surface reduces the collection efficiency and degrades the spatial resolution, as higher spatial frequency components of the evanescent field decay faster with distance.

An alternative approach is to replace the near-field probe with a focused beam of accelerated particles (electrons or ions).  Recently, focused electron beams have been used to excite optical modes in passive photonic and plasmonic structures, collecting emitted photons (cathodoluminescence[16,17]) or measuring the energy loss of the electron (electron energy loss spectroscopy[18–20]) to map the mode distribution.  A hybrid approach[21,22] uses femtosecond optical excitation in coincidence with energy-resolved ultrafast electron microscopy to probe the optically excited mode.  These techniques take advantage of the high spatial resolution of electron microscopy and avoid the limitations of the physical probe, but they have not demonstrated the necessary optical wavelength resolution to resolve the narrow high-Q modes that can be closely spaced in microresonators.

In this Letter, we demonstrate mapping of the electric field intensity distribution of whispering gallery modes in a silicon microdisk cavity (diameter = 10 µm, thickness = 245 nm) by detecting small perturbations to the excited optical mode induced by a focused lithium ion beam (FIB) probe.  While a lithium FIB scans across the microdisk resonator, an in-situ optical detection setup is employed to monitor the resonant wavelength of the optical mode. We measure a controllably small resonance shift which is closely related to the magnitude of the electric energy density at the location of the FIB probe. Among multiple possible interactions that can couple the FIB to the optical mode, we attribute the leading contribution to defect creation in the silicon lattice. Localized defects can mechanically deform the microdisk surface, shifting the resonance through an optomechanical perturbation. Comparisons with numerical simulations show that the measurements provide a faithful representation of the mode's spatial distribution with nanometer scale resolution.

A schematic of the experiment is shown in Fig. 1. A tunable diode laser is connected through a polarization controller to a fiber pigtailed on-chip waveguide[7] (Methods), which is evanescently coupled to the microdisk resonator. Optical transmission through the waveguide is measured by a high-speed photoreceiver. The absorption lines in the transmission spectrum in Fig. 2a indicate excitation of transverse magnetic (TM) polarized optical modes (electrical field oriented out of the plane of the microdisk). These modes can be grouped into families of the same radial order, each family having a distinct free spectral range and radial field distribution, which can be determined by numerical simulation (Methods) as shown in Fig. 2b. The optical setup allows tuning onto a selected absorption line and in-situ monitoring of the transmission while the device is placed in the vacuum chamber of a custom-built lithium FIB.

The lithium ion beam is generated by photoionizing ultracold lithium atoms held in a magneto-optical trap, accelerating the ions, and focusing them onto the sample[12] (Methods). The beam energy is kept at 3.9 keV, the instantaneous current is about 1 pA, and the beam is modulated by controlling the ionization laser with an acousto-optical modulator. We have chosen a lithium ion probe in part due to a unique combination of light mass and low penetration depth compared to other ion species available in FIBs (Supplementary Section 1). Compared to electrons, ions more readily produce lattice defects, which we expect to be the dominant interaction in our measurement.

To observe the effect of the FIB on the optical transmission, we first quickly scan the beam over a large area to locate and image the microdisk resonator by detecting the secondary electrons. This image serves as a map to selectively aim pulses of lithium ions at specific locations. Next, we set the monitoring laser wavelength (tunable between 1520 nm and 1570 nm) on the shoulder of an optical mode's absorption line (Figure 2c) and monitor the transmission change in response to short, rectangular ion pulses of controlled duration, typically 1 ms (Figure 2d). During the ion pulse, the observed transmission changes nearly linearly in time, while thereafter the accumulated change decays and settles to a finite residual value (Figure 2e). This transmission change only occurs when the ion beam is focused near the periphery of the microdisk, where the whispering gallery mode energy is localized. As Fig. 2c and 2d show, the change has opposite sign when the laser is fixed on either side of the resonance, indicating it is mainly due to a red shift of the resonance wavelength as opposed to a changing loss. With the ion beam focused near the peak of the $TM_{0,33}$ mode, we measure a resonant wavelength shift of approximately 1 pm ($\delta\lambda/\lambda \approx 7 \times 10^{-7}$) for a pulse containing 6000 lithium ions (Fig. 2d). Therefore, lithium ions are capable of generating a controllably small perturbation to the microdisk, with a perturbation strength related to the local optical mode intensity.

To measure the optical mode's spatial shape, we fix the laser wavelength on the blue shoulder of a given absorption line and monitor the optical transmission while the beam scans in a radial direction across the edge of the disk. Only radial line scans are performed because all the measured optical modes are traveling wave whispering gallery modes which are axially symmetric. The ion beam is modulated at 10 kHz and the change in transmission is measured via a lock-in amplifier. The resonant wavelength red shift is assumed to be linearly proportional to the measured transmission change, since the total shift for a single scan is typically less than 10 % of the resonance width. The results are shown in Fig. 3 for the zeroth, first, and second radial order TM modes. The mode shapes produced by lock-in measurement are similar to those produced by plotting the maximum transmission change during single pulse measurements (Supplementary Figure 2). For qualitative comparison, 2D profiles above the graphs in Fig. 3 depict the calculated electric energy density ($\varepsilon|\mathbf{E}^2|$) of the respective optical modes (Methods). A strong correspondence is evident between the data and the calculated energy density near the surface of the microdisk (the surface sensitivity is expected due to the shallow implantation depth of the low energy ions, about 30 nm), indicating the measurement provides a faithful representation of the mode distribution.

The mechanism by which a lithium ion beam causes a wavelength shift in the microdisk resonant modes can be elucidated by considering the various interactions that occur when an ion impacts the surface. These include multiple defect creations in the silicon lattice, sputtering, lithium implantation, heating, and charge deposition. As outlined in detail in Supplementary Section 2, we find that defect creation is likely to be the most important of these interactions, in part because each incident ion can generate a large number of vacancies (about 70 for our parameters, Fig. 1c). This results in a large amplification effect compared with other interactions.

Defects will perturb the optical resonance both by swelling the surface, thus changing the boundary geometry, and by changing the refractive index within the damaged volume. In Supplementary Section 2, we estimate the magnitude of these two perturbations as a function of ion dose and we use optical eigenmode perturbation theory[23] to calculate the expected wavelength shifts. We find that a defect-induced surface swelling, expected to be of order 100 pm, can by itself explain the observed wavelength shifts. The normalized results of the surface swelling perturbation theory are plotted as a function of position in Fig. 3 (orange solid lines), showing good agreement with the measurements. The optomechanical (surface swelling) perturbation is unique compared to any volumetric perturbation, such as a refractive index change, since the resonance shift depends differently on the in-plane and surface-normal components of the mode field (compare Supplementary Equations 2 and 4). This results in a distinct functional form of the expected curves, leading to smaller contrast between the signal peaks and nodes, as observed in the experimental data (Supplementary Figure 1). While a defect-induced refractive index change and other interaction mechanisms may also play a role, further experiments will be necessary to separately quantify each effect and develop a complete understanding of the underlying physics.

The combination of a FIB probe and in-situ optical detection opens up many interesting possibilities. A scanning beam probe is superior to a scanning tip probe not just for controlling the level of perturbation, but also because it enables imaging of microresonators with more complex three-dimensional geometries and mechanical sensitivity. Complicated nanopositioner feedback systems, required in NSOM to keep the probe at a constant height, are avoided. While our mode measurements can be done with controllably small damage, it is important to note that repeated probing with a FIB can eventually permanently spectrally shift and degrade the Q of the device. For example, after 12 radial scan measurements, the $TM_{0,33}$ mode was permanently red-shifted by roughly one linewidth, and its Q was reduced by about 15 % (Supplementary Section 4). This could be used as an advantage, for example, to control device uniformity post-fabrication, to alter the coupling strength in multi-microresonator systems[24], or to spectrally align, permanently and with great precision, the optical resonance with optical transitions in chip-scale cavity quantum electrodynamic systems[25].

Lastly, we note that it is possible to take advantage of our technique in a reverse sense, using the microresonator as a sensor to study the interaction of the ion beam with a solid. The response to single ion beam pulses (Fig. 2d and 2e) can be used as a real-time measure of defect creation and relaxation in the silicon lattice. Relaxation and diffusion of ion beam induced defects has been a problem of long-standing theoretical[13,26] and experimental[14,27] interest, primarily for the study of ion implantation in silicon microelectronics. Our methods may provide a way to study this relaxation with high spatial and temporal resolution and unprecedented sensitivity. The broadly successful application of microresonators as high-bandwidth and high-sensitivity detectors can now be extended to study the rich physics of ion-solid interactions.

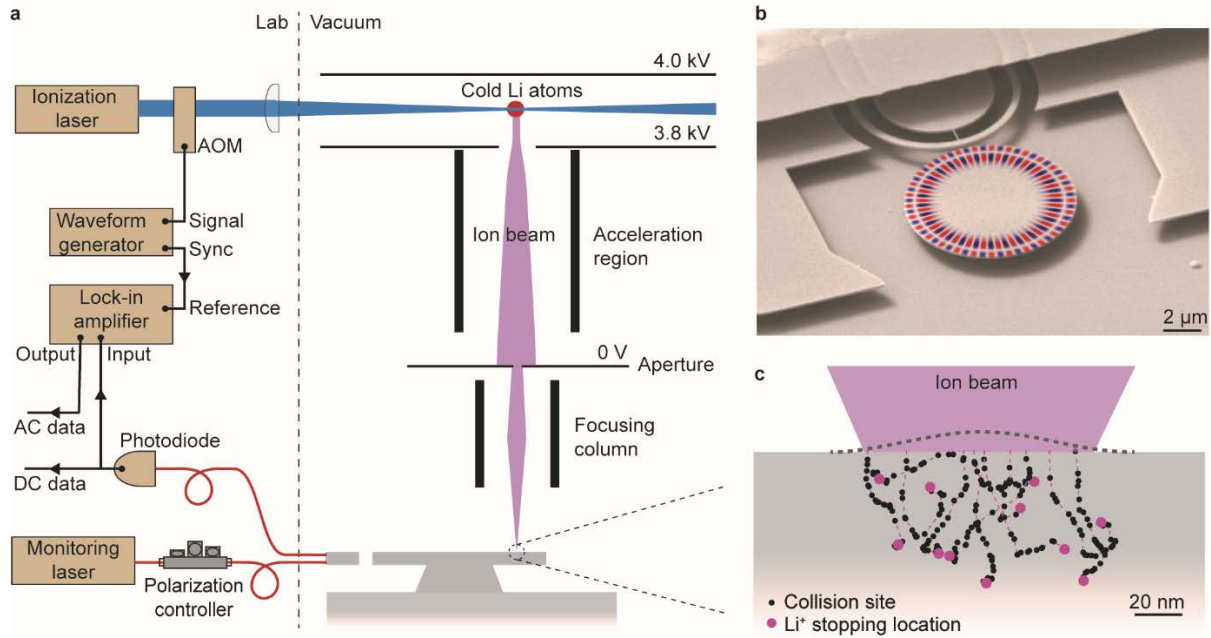

**Figure 1. Schematic of the measurement.** **a**, Light from a tunable diode laser (Monitoring Laser) is coupled into the microdisk through optical fibers and on-chip waveguides. A photodiode monitors the transmission of light at the output. The ion beam is produced by photoionizing cold lithium atoms held in a magneto-optical trap between two electrodes. The ions are accelerated and focused onto the microdisk. The ion beam is turned on and off by modulating the light from the ionization laser with an acousto-optic modulator (AOM). For dose-response measurements (DC data), a single pulse of ions is applied to the microdisk and the change in transmission is recorded directly from the photodiode as a function of time. For mode profile measurements (AC data), the ion beam is modulated at about 10 kHz and the photodiode output is sent through a lock-in amplifier to record the magnitude of the oscillation in optical transmission. **b**, Scanning electron micrograph of the microdisk cavity and coupling waveguide. The simulated mode profile of the $TM_{1,29}$ mode is superimposed. **c**, Schematic of lithium ion implantation at the microdisk surface. Example trajectories of $Li^+$ implantation are calculated (Methods) and displayed. Black circles mark collision sites where the $Li^+$ creates a vacancy or cluster of vacancies in the Si lattice, and purple circles mark the $Li^+$ stopping location. The localized defect creation can induce a surface swelling, indicated by the gray dashed line (the scale of the line is exaggerated for clarity).

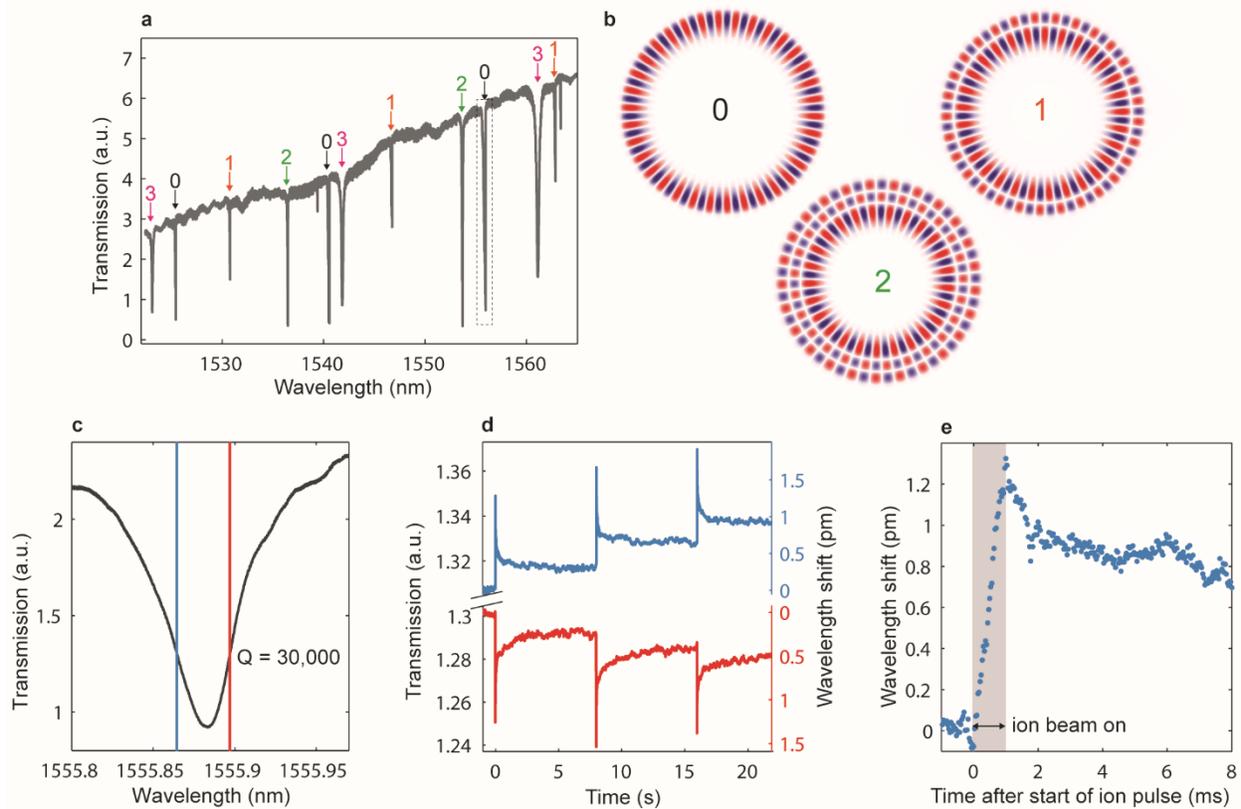

**Figure 2. Optical transmission through the microdisk cavity. a**, Measured optical transmission spectra. Sharp dips in the transmission indicate the wavelengths of high-Q cavity modes which have been labeled by their radial order number. **b**, Simulated 2D electric field distributions for the optical modes with different radial orders. **c**, Transmission spectrum for the single mode highlighted in **a**. For the ion beam measurements, the wavelength of the monitoring laser is fixed on either the blue or red side of the resonance (as indicated by the blue and red vertical lines) and the change in transmission is measured. **d**, Change in transmission in response to three ion beam pulses each of 1 ms duration applied at 0 s, 8 s, and 16 s. Blue and red lines show the response when the laser is tuned to the blue and red side of resonance. The absolute wavelength shift is calculated from a linear fit to the side of the resonance peak. **e**, Wavelength shift for the first pulse of the blue curve in **d** on a shorter time scale.

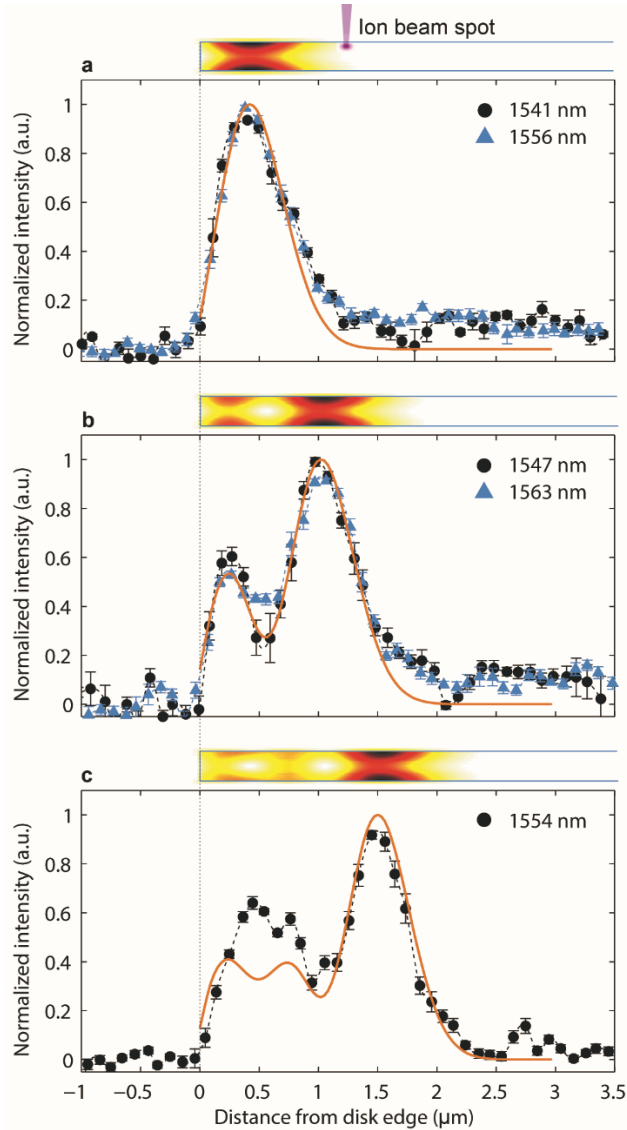

**Figure 3. Radial scan measurements compared to simulations for the zeroth (a), first (b), and second (c) order radial modes.** The 2D intensity maps show the simulated local electric field intensity for each mode. The black circles (and the blue triangles in **a** and **b**) are the normalized resonance wavelength shifts measured using the lock-in technique. The orange solid lines are the wavelength shifts calculated from the surface perturbation theory. Part **a** shows data for the zeroth radial order modes with two adjacent azimuthal mode numbers, $TM_{0,33}$ (1556 nm) and $TM_{0,34}$ (1541 nm). Part **b** shows data for the first radial order modes $TM_{1,29}$ (1547 nm) and $TM_{1,28}$ (1563 nm), and part **c** shows data for the second radial order mode $TM_{2,25}$ (1554 nm). The difference in the simulation results for adjacent azimuthal modes is within the thickness of the line (Supplementary Figure 1). Error bars represent one standard deviation statistical uncertainty of 4 to 15 measurements made at each ion beam position. The uncertainty in position is ± 50 nm, the approximate size of the ion beam. The ion beam size and depth profile is shown compared to the optical mode in **a**.

**METHODS**

**Device fabrication:** The silicon microdisk was fabricated on the device layer of a silicon-on-insulator (SOI) wafer. The oxide layer of the SOI wafer is 1 μm thick. Electron beam (E-beam) lithography followed by reactive-ion etching is performed to define the microdisk, waveguides, and light couplers. After removing the E-beam resist, layers of ≈ 2 μm silicon dioxide and ≈ 0.4 μm silicon nitride are then sequentially deposited. The nitride layer is photo-lithographically patterned and dry-etched to open windows for a later oxide removal step in the vicinity of the microdisk. Additional photolithography and etching are performed to produce the V-grooves that are aligned with the light couplers and waveguides. After removing etching masks and cleaving into individual chips, the device is dipped in a solution of 49 % hydrofluoric acid (HF) in water by weight. The HF removes the $SiO_2$ above and around the microdisk, suspending its edge and the portion of the on-chip waveguide evanescently coupled to the microdisk. Finally, cleaved optical fibers are glued onto the V-grooves to couple light in and out of the on-chip waveguides.

**Lithium FIB:** The lithium FIB is a custom FIB system that uses a new type of high-brightness ion source based on the photoionization of laser-cooled neutral lithium atoms. Lithium atoms are laser-cooled and trapped at a temperature of about 600 μK, and then photoionized in an electric field. The extracted ions are accelerated, coupled into a conventional ion column, and focused onto the sample[28]. Images are obtained by scanning the ion beam over the surface and collecting either secondary electrons (SE) or backscattered ions[12] similar to a scanning electron microscope. The system has demonstrated spot sizes of a few tens of nanometers at beam energies from 500 eV to 5 keV and beam currents of a few picoamperes. In this work, the operating ion energy is kept at 3.9 keV and the beam current is about 1 pA. The ion beam radius and the average implantation depth are about 50 nm and 30 nm, respectively. The spot size was measured by scanning the ion beam across the edge of the silicon wafer supporting the microdisk and measuring the change in SE signal. The implantation depth profile is calculated from Monte Carlo simulations using the SRIM software[29]. SRIM is also used to calculate the ion trajectories shown in Fig. 1c.

**Optical modeling:** Microdisk optical whispering gallery modes were computed with a full-vector finite element method. Time-harmonic solutions with spatial electric field distributions of the type $E_{n,m}(r,z)\exp(i*m\varphi)$ were obtained (r, φ and z are the radial, azimuthal and axial coordinates), where m and n are the azimuthal and radial modal order numbers. For a fixed azimuthal order m (m = 0, 1, 2, …), a variety of modes with radial order n and eigenfrequency $\omega_{n,m}$ were obtained. The refractive index of silicon was considered to be $n_{Si}$ = 3.48, and the disk radius and thickness were varied within reason until the computed eigenfrequencies approximately matched the measured optical transmission spectrum. This allowed us to assign each transmission dip to one mode of radial order n, as shown in Fig. 2a, all assigned to transverse magnetic modes, $TM_{n,m}$, with major magnetic field component parallel to the disk plane. The roughness of the sidewalls of the microdisk, which can reduce the Q of the resonances, was not considered in the numerical simulation. According to scanning electron micrographs, the sidewall roughness is tens of nanometers at most and considerably smaller than the wavelength. We expect that this omission causes negligible perturbations (i.e., much smaller than our measurement error) to the calculated mode shapes.

**ACKNOWLEDGEMENTS**

K.A.T. and J.Z. acknowledge support under the Cooperative Research Agreement between the University of Maryland and the National Institute of Standards and Technology Center for Nanoscale Science and Technology, Award 70NANB10H193, through the University of Maryland.

**AUTHOR CONTRIBUTIONS**

K.A.T, J.Z., J.J.M. and V.A.A. designed the experiments.  J.Z. and V.A.A. fabricated and characterized the microdisk.  K.A.T. and J.J.M. operated the lithium ion beam instrument.  K.A.T. and J.Z. analyzed the data and wrote the draft manuscript.  M.D. and K.S. performed the optical modeling and mode perturbation calculations.  All authors contributed to interpreting the data and editing the manuscript.

**COMPETING FINANCIAL INTERESTS**

The authors declare no competing financial interests.

# Imaging Nanophotonic Modes of Microresonators using a Focused Ion Beam

Kevin A. Twedt[a,b]†, Jie Zou[a,b]†, Marcelo Davanco[a], Kartik Srinivasan[a], Jabez J. McClelland[a], Vladimir A. Aksyuk[a]*

[a]Center for Nanoscale Science and Technology, National Institute of Standards and Technology, Gaithersburg, MD 20899, USA
[b]Maryland Nanocenter, University of Maryland, College Park, MD 20742, USA

†These authors contributed equally to this work.
*vladimir.aksyuk@nist.gov

## Supplementary Information

**Supplementary Discussion**

**S1. Advantage of the lithium ion source**

The lithium ion source was chosen in part due to the unique combination of light ionic mass and low operating energy (typically 500 eV to 5 keV). The light mass ensures that our probe will cause a minimum of damage to the microdisk, and the low energy ensures a low penetration depth, making the probe more surface sensitive. Lithium is also known to insert interstitially and readily diffuse in silicon, and thus the cumulative damage from repeated exposure is likely to be smaller than for less mobile ions. Examples of other ions species that are readily available as focused ion beams (FIB) include gallium and helium. Gallium has a shallow implantation depth, even at higher energies than that of our lithium beam, but has a sputtering rate typically ten times higher than lithium, and does considerably more damage to the target at the same beam current. Helium has a lower sputter rate than lithium, but is available only at higher beam energies (> 10 keV) where the penetration depth would be hundreds of nanometers. Many of the He ions would travel through the full thickness of the microdisk, complicating the interpretation of the mode measurement. The cold-atom-based lithium FIB also allows easy and accurate time control of the beam pulse. We can modulate the current of the ion beam at the source simply by modulating the intensity of the ionization laser with an acousto-optic modulator, which allows arbitrary modulation at speeds up to 50 MHz. For all of these reasons, we found the lithium FIB to be a good choice for demonstrating this mode imaging technique. It is possible that other ionic species, or possibly even electrons, may work as well; however further experiments will be needed to determine whether this is so.

**S2. Explanations for a FIB-induced resonance shift**

*S2.1 Interaction of the ion beam with the target*

When an ion enters a target, it loses energy through various interactions with both the atoms and the electrons in the material and eventually comes to rest. This process results in a number of changes to the target, including the creation of defects in the crystalline lattice, sputtering of atoms from the surface, implantation of the incident ions, local charge deposition, and heating. Various theoretical models and computer simulations have been developed to facilitate understanding of the relative importance of all these effects as a function of ion beam parameters. Here, we use the popular Monte Carlo SRIM software[1] to calculate the size of the interaction volume and to compare the relative magnitudes of structural changes to the target.

First we calculate the interaction volume and the distribution of implanted lithium ions. For a $^7$Li ion beam of 3.9 keV energy incident on a silicon target, the average implantation depth (range) is 30 nm and the lateral straggle (assuming a point implantation spot on the surface) is well fit by a Gaussian with a 20 nm standard deviation. The 20 nm straggle is smaller than the nominal 50 nm radius of the spot size from the lithium FIB, so this adds minimally to the lateral size of the implantation volume. The precise depth profile of implanted ions is determined from a fit to the result of 40,000 ion trajectory simulations, allowing us to write down an expression for the density of implanted lithium assuming a Gaussian ion beam with standard deviation of 50 nm:

$$n(x,y,z) \propto \left(\frac{z-z_1}{\sigma_z}\right)\exp\left(-\frac{(z-z_2)^2}{2\sigma_z^2}\right)\exp\left(-\frac{x^2}{2\sigma_x^2}\right)\exp\left(-\frac{y^2}{2\sigma_y^2}\right) \qquad (S1)$$

with $z_1$ = -16.9 nm, $z_2$ = 21.1 nm, $\sigma_z$ = 18.8 nm, and $\sigma_x = \sigma_y$ = 54 nm. In Eq. S1, the x and y coordinates are in the plane of the microdisk, x = y = 0 is the center of the beam focal spot, and z = 0 is at the surface with positive values going into the resonator.

While Eq. S1 is the expected lithium distribution immediately following implantation, we note that in general there will be a certain amount of diffusion over time. Likewise, the energy and charge delivered by the ion beam, as well as the damage created, will initially be localized in this volume, but can also diffuse outward. In the following discussion, we will address the importance of diffusion for each of the physical processes that may be influencing the measurement. Provided our measurement time is short compared with the time constants for diffusion out of the implantation volume, however, good lateral resolution and surface sensitivity will be maintained in the measurement.

Along with the spatial distribution of implanted lithium, the simulations also provide information on sputtering and vacancy creation in the silicon lattice. For 3.9 keV Li ions in Si, we find that the most significant target change is the creation of vacancies. At this energy, Li ions create an average of 70 vacancies per incident ion, whereas they sputter at a rate of only 0.4 silicon atoms per incident ion, and implant at a rate of only 0.92 lithium atoms per incident ion, the remaining 8 % of incident ions being backscattered. Note that the SRIM software considers only the creation of simple vacancies, though more complicated defects may also form, and does not consider any relaxation or diffusion of the defects which may occur after their creation.

The large ratio of vacancies created to implanted atoms gives any interaction mechanism associated with lattice defects a large amplifying factor. Thus we will begin, in sections S2.2 through S2.4, by discussing the ways in which defects can lead to a shift in the optical resonance. Then, in section S2.5, we return to consider the effects of other possible interaction mechanisms.

*S2.2 Interaction of FIB-induced defects with the optical mode*

The creation of a local concentration of defects in a microdisk resonator has the possibility of coupling to the optical mode and altering the resonant wavelength in two distinct ways. First, the geometry of the resonator surface may be locally altered. Ion beams in low doses can induce a local surface swelling as a direct consequence of the large number of defects created.[2,3] With higher doses of heavy ions (e.g. gallium), the size of this swelling quickly saturates and then reverses as the bombarded region becomes amorphous and sputtering takes over. In our case, the ion doses are low enough and the sputtering poor enough that milling should not occur and swelling should dominate. Second, the refractive index may be locally altered since the defect creation partially turns the crystal amorphous and decreases its density. While both of these effects are likely to happen and will alter the resonance at some level, for simplicity we will consider them separately.

It is difficult to precisely calculate the magnitude of the surface swelling and refractive index change that a given ion dose will produce. Therefore, we instead start with the measured wavelength shift and work backwards to determine whether this shift is consistent with the number of defects created by the given ion beam dose. Throughout the following discussion, we refer to the measurement in Fig. 2d and 2e, where a 1 ms pulse of ions at 1 pA beam current is placed at the maximum of the $TM_{0,33}$ mode. The lithium ions are implanted into the volume given by Eq. S1. The resulting lithium density at the maximum of the distribution is 7 x $10^{24}$ $m^{-3}$ and the predicted initial vacancy density is 5 x $10^{26}$ $m^{-3}$, or about 1 vacancy for every 100 silicon lattice sites. This dose produces a transient optical resonance frequency (vacuum wavelength) red shift of about 0.1 GHz (1 pm).

S2.2.1 Surface swelling perturbation

We use optical eigenmode perturbation theory to calculate the amount of surface swelling needed to shift the resonant wavelength by 1 pm, and then compare this calculated swelling to an estimate based on the given ion dose. Wavelength shifts of the whispering gallery modes due to local swelling of the microdisk can be estimated by modeling the effect of a small local deformation of the top microdisk surface, where the ion beam is incident. The following perturbative expression can be used[4]

$$\frac{\delta\lambda}{\lambda} = -\frac{\delta\omega}{\omega} \approx \frac{1}{2}\frac{\int |Z|[\delta\varepsilon |\mathbf{E}_\parallel|^2 - \delta(\varepsilon^{-1})|\mathbf{D}_\perp|^2]dA}{\int \varepsilon|E|^2 dV} \quad (S2)$$

Here, $\lambda$, $\omega$, **E**, and **D** are the wavelength, frequency, modal electric field and electric displacement field of the unperturbed disk, respectively, $\delta\varepsilon$ = $\varepsilon_{Si} - \varepsilon_{air}$, $\delta(\varepsilon^{-1})$ = $1/\varepsilon_{Si} - 1/\varepsilon_{air}$, and $\varepsilon_{Si,air}$ are the permittivities of silicon and air. The swelling is represented by an outward boundary displacement Z normal to the top disk surface. We assume that the spatial dependence of Z is the same as given in Eq. S1 for the lateral profile of the incident beam,

$$Z(x,y) = \frac{Z_{\max}}{2\pi\sigma^2} \cdot \exp\left[\frac{-(x-x_0)^2-(y-y_0)^2}{2\sigma^2}\right], \quad (S3)$$

where $(x_0,y_0)$ is the position of the center of the ion beam on the disk surface and $\sigma$ = 54 nm. In Eq. S2, the integral in the numerator is performed over the entire surface of the nanostructure, while that in the denominator is performed over the entire space. To compare to the example measurement, we evaluate Eq. S2 for the $TM_{0,33}$ mode at a position $(x_0,y_0)$ where the electric field intensity is maximum. We find that a maximum surface displacement of $Z_{\max}$ = 0.4 nm is needed to produce a 1 pm red shift of the mode.

A simple upper estimate for the amount of surface swelling from a given ion dose can be calculated by assuming each created vacancy (predicted by SRIM) survives and adds vertically to the silicon lattice size. This predicts a maximum surface swelling of about 0.5 nm for the above parameters. Several previous experiments have also measured an ion-induced surface swelling of silicon[3,5] and other targets[2], typically with ex-situ atomic force microscopy. Since these experiments use different ion species, beam energies, and total fluences, all of which affect the amount of surface swelling, it is difficult to draw any direct comparison to our work or even to assess the validity of the simple upper estimate. Still, it is plausible that our lithium ion dose is capable of producing a localized surface swelling large enough to explain the observed wavelength shift.

S2.2.2 Refractive index perturbation

Wavelength shifts of the whispering gallery modes due to perturbations of the local refractive index induced by the ion beam can be obtained with the first-order perturbation theory expression[4]

$$\frac{\delta\lambda}{\lambda} = -\frac{\delta\omega}{\omega} \approx \frac{1}{2}\frac{\int \delta\varepsilon |\mathbf{E}|^2 dV}{\int \varepsilon |\mathbf{E}|^2 dV}, \quad (S4)$$

where the integrals are taken over the entire space. In Eq. S4, $\omega$ and $\mathbf{E}$ are the whispering gallery mode frequency and electric field profile for the unperturbed disk, $\varepsilon$ is the spatially-varying dielectric constant, and $\delta\varepsilon$ is the dielectric constant perturbation. Assuming that the perturbation $\delta\varepsilon$ follows the shape of the lithium implantation profile in Eq. S1, we write

$$\delta\varepsilon(x,y,z) = \frac{(\delta\varepsilon)_{\max}}{N} \cdot \left(\frac{z-z_1}{\sigma_z}\right) \cdot \exp\left[-\frac{(z-z_2)^2}{2\sigma_z^2}\right] \exp\left[\frac{-(x-x_0)^2-(y-y_0)^2}{2\sigma^2}\right], \quad (S5)$$

where $(x_0,y_0)$ is the position of the center of the ion beam on the disk surface, and N is a normalization factor. We substitute Eq. S5 in Eq. S4 and again evaluate for the $TM_{0,33}$ mode at a position $(x_0,y_0)$ where the field intensity is maximum. We find that a fractional increase in the dielectric constant of $(\delta\varepsilon)_{\max}/\varepsilon$ = 3.2 % is needed to produce a 1 pm red shift of the mode. This corresponds to an increase in the refractive index of 1.6 %, making use of the relation $\delta\varepsilon/\varepsilon = 2\delta n/n$.

Defect creation by the ion beam can change the refractive index because it partially turns the crystal amorphous and decreases its density. Amorphous silicon created by ion bombardment can have a 10 % to 15 % larger refractive index than crystalline silicon[6] so the direction of the shift is consistent with our measurement. Since we generate at most 1 vacancy for every 100 lattice sites, the expected change in refractive index due to amorphization is only 0.1 %, an order of magnitude too small. The refractive index may also change due to the strain applied to the lattice by the increased volume. Based on a surface swelling of order 0.1 nm and an implantation depth of order 10 nm, a strain of order 1 % may be a reasonable estimate, although the unusual spatial profile makes it difficult to interpret. One previous study[7] found that a 1 % strain in a silicon microstructure can induce a refractive index increase of 0.5 %, large enough that this effect cannot be ruled out.

*S2.3 Relaxation of FIB-induced defects*

From the previous section, we can conclude that FIB-induced defects are capable of generating a transient wavelength shift consistent with our measurements. It is also useful to discuss whether defects can be responsible for the observed relaxation in the wavelength shift. Figure 2d and 2e show an example of a single pulse implantation measurement plotted on two time scales. The decay in the signal begins immediately after the ion beam is turned off and can persist for several seconds. This type of extended relaxation appears to be consistent with at least one recent theory[8], but despite decades of research[9–11] there is not a clear consensus on a value for the time constant of defect relaxation in silicon, and the ion beam energy, flux, and fluence used are likely to have an influence on how the system relaxes. With further study, our measurements could prove to be a powerful new way to study defect relaxation in real time at speeds faster than have previously been accessible to experiments.

*S2.4 Spatial variation of resonance shifts*

For both the surface swelling and the refractive index change, the theory can be applied to generate the expected wavelength shifts as a function of ion beam position across the disk surface. We perform this calculation for varied radial position for each of the five measured modes and plot the results in Fig. S1. The normalized surface swelling curves are also shown in Fig. 3. The results for adjacent azimuthal modes are nearly identical in both the calculation and in the data, as expected for these high azimuthal order numbers.

Consider the shape of the theory curves in Fig. S1. The surface swelling calculation has less contrast between the peaks and valleys in the first and second radial order modes (Figs S1c-f), due to the asymmetric interaction with the electric field components at the surface. This feature is unique to the optomechanical perturbation. Any perturbation that is local in volume only (i.e. does not move the surface) will interact with all vector components of the electric field equally, as in Eq. S4, and will have a

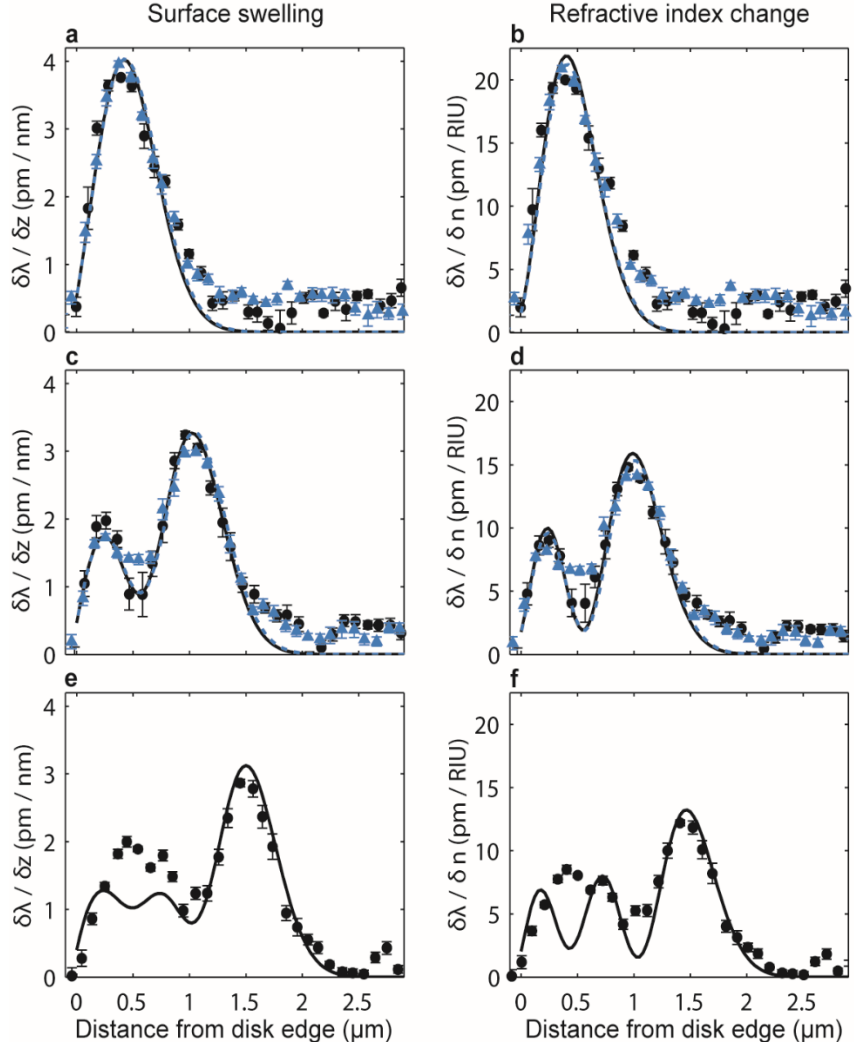

**Figure S1. Calculated radial mode profiles. a,c,e,** Mode profiles calculated using the surface swelling perturbation theory compared to data for the $TM_{0,34}$ (**a**, black solid line and black circles, 1541 nm), $TM_{0,33}$ (**a**, blue dashed line and blue triangles, 1556 nm), $TM_{1,29}$ (**c**, black solid line and black circles, 1547 nm), $TM_{1,28}$ (**c**, blue dashed line and blue triangles, 1563 nm), and $TM_{2,25}$ (**e**, black solid line and black circles, 1554 nm) modes. The units are wavelength shift (in pm) divided by the height of the swelling at the maximum of the distribution (in nm). **b,d,f,** Mode profiles calculated using the refractive index perturbation theory compared to data for the same modes shown in **a,c,e** respectively. The units are wavelength shift (in pm) divided by the change in refractive index (refractive index units). The data and surface swelling calculations are the same as was presented in Fig. 3. Each data set is scaled to the corresponding theory curve for each mode individually.

radial profile exactly the same as in the right column of Fig. S1, whether the perturbation is to the refractive index or some other quantity. An altered surface geometry is the only perturbation that interacts with normal and tangential components differently, as in Eq. S2. Comparing the two sets of theory curves to our measurements, we find that the uncertainty in the data is too large to clearly distinguish between the two theories for the zero or first radial order modes. For the second radial order mode, the measured curve has only slight contrast between the peaks and valleys, more closely
19

matching the surface swelling calculation. This adds to the evidence in support of defect-induced surface swelling as the dominant mechanism.

Note that all of the theoretical mode shapes presented in this paper are calculated independently of the FIB measurements, as described in the Methods and section S2.2. However, the theory curves are scaled in magnitude to provide a straightforward comparison with the measurements. In Fig. 3, the theory curves are scaled in the vertical axis and offset in the horizontal axis to fit to each of the five data curves individually. Then the data and theory curves are both normalized to put the theory curve on a 0-to-1 scale. In Fig S1, the same procedure is used, but the data curves are normalized to the calculated magnitude of each theory curve. We present comparisons of normalized theory and data curves to highlight the qualitative differences in the shapes. There are too many uncertainties in the calibration of the absolute magnitude to present a precise quantitative comparison. However, it is also important to note that the measured magnitude of the wavelength shifts for all modes are of the same order, consistent with the theoretical calculations. For data taken with approximately the same parameters, the measured relative magnitudes of the first order and second order radial modes are the same, to within 10 %, in agreement with the theory. The measured peak magnitude of the zero order radial mode is about 80 % larger, compared to just 30 % larger in the theory. It is possible that this discrepancy is due to systematic effects such as a drift in the absolute ion beam current between measurements.

*S2.5 Other possible explanations for the resonance shifts*

While defect creation appears to be a promising candidate to explain the resonance shifts, we also consider possible effects from heating and lithium ion implantation. An increase in temperature from the energy deposited by the ion beam will change the refractive index, but the change is only localized for times shorter than the thermal diffusion time. The thermal conductivity of silicon is large enough that the time to diffuse out of the 1 μm mode length should be on the order of microseconds, much faster than the rise and decay in the measurement. If heating were the dominant interaction, we might observe a wavelength shift, but we would not expect to observe any spatial variation.

Lithium implantation into silicon at high doses can result in various alloy phase changes[12] that certainly alter the dielectric properties, but the lithium dose in our measurement is so low (a maximum Li/Si ratio of about $10^{-4}$ in the 1 ms pulse measurement) that these phase changes are not likely to occur and it is hard to imagine a 1.6 % change in refractive index. Additionally, if lithium doping were the dominant interaction, then the observed relaxation of our signal would presumably be due to lithium diffusion out of the optical mode volume. Diffusion constants for lithium in silicon vary[13,14], but even for the large value of $10^{-13}$ m$^2$/s, diffusion to a 1 μm length scale would take 10 s, slower than the relaxation we observe.

*S2.6 Conclusion*

We have discussed multiple possibilities to explain how focused ion implantation can alter the properties of a microresonator, couple to the optical mode, and shift the resonant frequency. Defect creation by the ion beam is the only interaction mechanism we can find that is capable of generating a wavelength shift that is local in space, with a magnitude and relaxation time consistent with our measurements. Furthermore, we have identified two ways in which defect creation can couple to the optical mode, through a surface swelling or through a refractive index change. Based on the magnitude of the effect and the shape of the radial profile measurements, surface swelling is probably the larger of the two effects. In reality, though, the surface swelling and refractive index change are intimately coupled, so both these effects and possibly other effects need to be considered together to develop a complete picture of the optical mode perturbation. Experiments using other ion species and with other dielectric materials will be necessary to confirm our interpretation and gain a better understanding of the underlying physics.

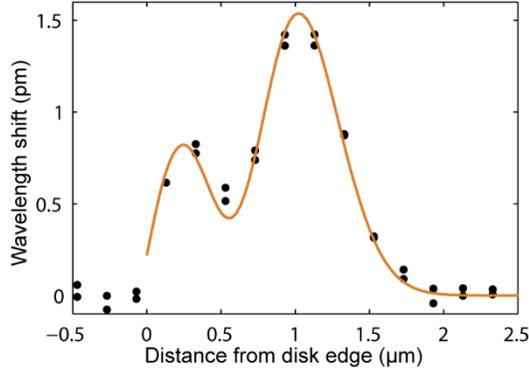

**Figure S2. Radial profile measurement with single pulse implantation.** Measured total wavelength shift (calculated from transmission signal) of the $TM_{1,29}$ mode in response to a 1 ms ion pulse applied at varied position. The black data points are from two independent measurements made at each position, and the orange line is the normalized theory (same as in Fig. 3b), scaled to fit the data.

### S3. Radial profile measurement with single pulse implantation

In addition to the radial profile measurements made with the lock-in technique and shown in Fig. 3, we have also measured the radial profile with a series of short pulse measurements, similar to Fig. 2e, taken at varied position. This data is presented in Fig. S2 for the $TM_{1,29}$ mode. The measured radial profile shape is similar to that presented in Fig. 3 and again consistent with the surface swelling theory. We do not notice any systematic variation in the pulse temporal response as a function of position, other than the overall magnitude.

### S4. Cumulative impact of measurements on device resonance frequency and quality factor

It is important to note that mode shapes can be obtained with minimal invasion, despite the inherently destructive interaction of an ion beam with its target. As was demonstrated in Figs. 2d and 2e, short ion pulses have a negligible and mostly reversible effect on the optical resonance. We can easily measure wavelength shifts of a fraction of 1 pm, less than 1 % of the resonance width and limited only by the signal-to-noise ratio (SNR) of the optical detection. Likewise, a single radial line scan measurement with a sufficient SNR to identify the basic mode shape deposits only $10^5$ ions (1 pA for 18 ms) and results in no measurable reduction in Q (< 1.2 %).

Eventually, repeated probing of a device with an ion beam will destroy the device either through sputtering, or in our case, dosing the disk with a large amount of lithium. However, we found that after several days of experiments on one device, including all the data presented in this paper, the Q degraded by only a factor of 2.5, from Q = 46,000 to Q = 18,000. During this time, the disk received a dose of greater than $10^7$ lithium ions. If these ions were to diffuse uniformly through the disk volume, the lithium density would be $10^{24}$ $m^{-3}$. If the diffusion is less significant, the density near the measurement locations at the edge of the disk may be much higher. Apparently, this amount of lithium doping does not meaningfully impact the optical properties, which may not be true if the measurement were performed with a different ion species, e.g. gallium.

The measured amount of Q degradation is by no means the fundamental limit for the technique For example, we expect that simple improvements to device fabrication can reduce the optical coupling losses by a factor of 10 or more, improving the SNR of the measurement and reducing the necessary ion dose.  Additionally, performing the measurements at elevated temperature may make the recovery of the FIB-induced damage (Fig. 2d) more complete, reducing the permanent changes to the device.  These and other refinements to the technique should further reduce the amount of damage needed to make useful mode measurements.

The permanent, cumulative changes to the device could alternatively be used as an advantage, as a way to precisely trim the resonance wavelength or quality factor to match a desired value.  Precise trimming of cavity resonances is useful in cavity quantum electrodynamics in order to spectrally align a cavity with narrow line emitters such as quantum dots.  In our case, the ion beam causes both a red shift and a broadening of the spectral line, so the ability to trim the resonance will be limited by the relative size of these two effects.  For example, after 12 radial scan measurements (each with a typical dose of $10^5$ ions), the $TM_{0,33}$ mode was permanently red-shifted by about one linewidth, and its Q was reduced by about 15 %, from Q = 46,000 to Q = 39,000.  The ratio of line shift to line broadening is roughly 7 to 1.  This suggests the ability to tune the resonance wavelength by a few linewidths, sufficient to address multiple spectrally distinct states.

The mode imaging technique can in principle be extended to measure microresonator modes with arbitrarily high Q.  For a higher Q device, the optical transmission measurement is more sensitive to shifts in the resonance, so the mode measurement can be made with a reduced ion dose and thus reduced damage.  If the line shift and the line broadening scale with the ion dose in a similar way, the measurement will degrade the Q of the device by the same percentage, regardless of the initial value of Q.  This is a primary advantage of using a charged particle probe: the probe strength is easily varied to match the sensitivity of the device.

Finally, it is important to note that calculated mode shapes do not vary significantly with Q as long as Q remains relatively high (>10,000).  The observed degradation in Q is most likely due to additional roughness induced by the ion impact, which leads to greater scattering. This roughness is much smaller than the wavelength and hence has negligible impact on the mode shape.  Thus the gradual reduction in Q observed in the experiment should not significantly affect the mode shape measurements.